# Calculated Pre-exponential Factors and Energetics for Adatom Hopping on Terraces and Steps of Cu(100) and Cu(110)

Handan Yildirim[1,2], Abdelkader Kara[1,*], Sondan Durukanoglu[2], and Talat S. Rahman[1]

[1]Department of Physics, Kansas State University, Manhattan KS 66502, USA
[2]Department of Physics, Istanbul Technical University, Istanbul, Turkey

**Abstract**

We have calculated the vibrational dynamics and thermodynamics for Cu adatom hopping on terraces and near step edges on Cu(100) and Cu(110), using the embedded atom method for the interatomic potential. The local vibrational densities of states were calculated using real space Green's function formalism and the thermodynamical functions were evaluated in the harmonic approximation. The calculated diffusion energy barriers for six specific local environments on Cu(100) agree well with experimental and previous theoretical results. Contribution of vibrational entropy to the change in the free energy of the system as the adatom moves from the equilibrium configuration (hollow site) to the saddle point, is found to be as much as 55meV (144 meV) at 300K (600K). The prefactors for all 13 cases are found to be of the order of $10^{-3}$ cm$^2$/s, almost independent of temperature, and the respective activation energy barriers.

[*]Corresponding author: akara@ksu.edu; tel.: 17855325520.

## I. Introduction

Understanding surface diffusion processes is of vital importance to studies of surface related phenomena such as crystal growth, thin film growth, surface chemical reactions and catalysis. However, in spite of the conceptual simplicity, the phenomenon of diffusion of adatoms or small clusters on clean, infinite, defect-free surfaces stays a challenging problem. The first surface diffusion study using field ion microscopy (FIM) was reported by Muller [1], and since then, surface diffusion of a single atom on various metals has been observed [2,3], as FIM is capable of resolving individual atoms, although observations have been limited to a few metals: tungsten [4,5], rhodium [6], platinum [7,8], nickel [9], and iridium [10,7,3]. From the usually observed Arrhenius form of the diffusion coefficient (D), the diffusion pre-exponential factor ($D_0$), also called the prefactor, and the activation energy barrier $\Delta E$ are generally extracted using:

$$D = D_0(T) \exp\left(\frac{-\Delta E}{k_B T}\right) \qquad (1)$$

Even after several instrumental advances, reliable diffusion data are available only for the simplest processes on a small number of simple surfaces [11]. Because diffusion is an activated process, small errors in energy barriers translate into large uncertainties in the diffusion coefficients [12]. To determine the pre-exponential factor from observed Arrhenius behavior, several measurements are needed in a reasonable range of temperature. Since such measurements are difficult, quite often the prefactor is simply assumed to be close to a so-called 'usual value' of about $10^{-3}$ cm$^2$/s, although experimental results have quoted "abnormal" values for the prefactor [13]. While most theoretical studies of surface diffusion also ignore explicit calculations of diffusion prefactors, some attemps have already been made to recognize the significance of vibrational entropy contributions [14,15], in calculations of diffusion prefactors. Of particular relevance to our work here are a series of papers in which diffusion prefactors have been calculated with explicit inclusion of the vibrational dynamics of the system with the moving entity at both the equilibrium position and at the saddle point. In one set of calculations such prefactors are calculated using Vineyard's formula [16]

$$D_0 = \prod_{i=1}^{N} \nu_i / \prod_{j=1}^{N-1} \nu_j \qquad (2)$$

Where the $\nu_i$'s and $\nu_j$'s are the set of vibrational frequencies of the system for the equilibrium and saddle point configurations, respectively. In the other set of calculations, Kurpick and co-workers [17-20] have developed a recipe for calculations of the prefactors through evaluations of the changes in the vibrational contributions to the free energy of the system, within the limits of validity of the transition state theory [21].

The goal of this study is to examine the influence of the local environment on the prefactor for the diffusion of a single atom on metal surfaces. For this purpose, using the calculational scheme developed by Kurpick *et al* [17-20], we have carried out a systematic study of the activation energy barriers and the corresponding prefactors for the diffusion of a Cu adatom on the terraces and at and near the step edges of Cu(100) and Cu(110). An interesting feature of prefactors was introduced by Meyer-Neldel [22] who suggested that for processes where the activation energy is larger than both the energies of the activation and $k_BT$, there is a compensation by which the prefactor increases exponentially if the activation energy increases [23]. The rule has been tested by means of molecular-dynamics simulations for adatom diffusion on and near Cu(100) surface, and shown to remain valid [24]. On the other hand similar studies for Cu and Ni surfaces using semi-empirical potentials show that the compensation rule is not always upheld [25]. With the set of activation energy barriers and prefactors that we calculate, we also intend to check the applicability of the Meyer-Neldel rule. It should be pointed out that a few of the results that we present here, for example that of an adatom on Cu(100), are already available in the literature on the subject. Wherever possible we refer to available results and include ours for completeness and comparison. Note also that in this work we have confined ourselves to diffusion via hoping. While exchange mechanism is possible, preliminary investigations and previous work show the barrier for exchange, for several of the cases presented here, to be larger than that for hoping, For example, while for adatom hopping process on Cu(100) the barrier is found to be 0.53 eV, for the exchange process it is 0.79 eV [26]. The barriers in case of Cu(110) surface diffusion along the open channel are 0.24 eV for hopping and 0.87 eV for the exchange processes [27]. On the other hand the activation energy barrier for adatom exchange at a step edge (diffusion over the step) on Cu(100) is found to be 0.51 eV, while that for hopping is 0.77 eV [26]. In such a case exchange mechanism is more likely. Nevertheless for purposes of comparison of the prefactor for a set of adatom hopping processes, the role of exchange mechanism in diffusion has been set aside for future consideration.

The paper is organized as follows: In Section 2, we present our system of interest. Section 3 includes the theoretical details which are applied during the calculations containing three subsections. We present and discuss our results in Section 4 and finally we conclude in Section 5.

**II. Systems of Interest**

We present in figures 1 and 2 several processes involving hopping of an adatom from an fcc hollow site to a neighboring hollow site on the terraces and near step edges of fcc(100) and (110) surfaces. Figure 1 shows the minimum-energy configuration in which an adatom is adsorbed on a hollow (four-fold site) near or at a step edge or far from it. The arrows show the direction along which the adatom would perform the process labeled in the figure. The process labeled (P1) in Fig. 1 corresponds to a hop on the (100) terrace, while processes (P2), (P3) and (P4), associated with an adatom originally on the upper terrace and at the step edge, correspond to a jump away from the step edge, decent from the step and along the step, respectively. Finally, processes (P5) and (P6) correspond to diffusion, on the lower terrace, away from and along the step edge, as shown in the figure. In figure 2, we present the processes studied here for an adatom hopping on an fcc(110) terrace and at near a step edge on this surface. The fcc(110) surface may be envisioned as an arrangement of dense chains running parallel to each other and separated by a distance equal to the lattice constant of the element. An atom adsorbed on the hollow 4-fold site may then diffuse parallel to these chains (open channel), shown in Fig. 2 as process (P7), or perpendicular to the open channel, labeled process (P8). When the adatom is placed on the upper terrace and at the step edge, it may diffuse along the step edge (P11), away from it (P10), or jump to the lower terrace (P9). The processes labeled (P12) and (P13), represent diffusion of an adatom away from and along the step edge, on the lower terrace.

**III. Theoretical Details**

In this section, we discuss some details of the theoretical techniques which are used to calculate the structure, energetics, vibrational densities of states, thermodynamics, prefactors and diffusion coefficients. These are presented in three subsections representing summaries of calculational details for: i) activation energy barriers, ii) local vibrational densities of states (LDOS) iii) thermodynamic functions, prefactors and diffusion coefficients. These details have already been presented in several publications [17-20] and are included here only for completeness.

**III.1. Activation Energy Barriers**

To determine the static energy barriers we perform a series of energy-minimization, 'molecular statics' (MS) calculations. In order to obtain the relaxed configurations, standard conjugate gradient method is used for minimizing the total energy of the system [28]. For these calculations, the important ingredient is the interaction potential between the atoms. In our case we use semi-empirical many body potentials as obtained from the embedded atom method (EAM) [29]. When the position of the adatom at the saddle point is known (by symmetry, for example), only two calculations are performed, one when the adatom is at the minimum energy site and the other when the adatom is put at the saddle point. When adatom diffusion occurs near a complex geometry (involving a step, for example), the atomic configuration of the system at the saddle point is not known 'a *priori*' and depending on the complexity of the system, one may use sophisticated methods like the nudged elastic band [30], the drag or grid methods [31] to calculate the activation energy barriers. Here we perform a 1D scan of the energy landscape between positions of two consecutive minima of the adatom.

## III.2. Local Vibrational Densities of States

There are several theoretical techniques for calculating the vibrational densities of states. The mostly used one is the slab method in which one needs to diagonalize the dynamical matrix portraying the force constants between the particles in N layers of the slab [32]. The continued fraction (CF) method using real space Green function is another way to calculate the vibrational dynamics of surfaces [33]. Since, our interest lies in obtaining the local contributions to the dynamics and thermodynamics of systems which have site specific environments, a local approach in real space is more appropriate than the one based on k-space. The real space Green function (RSGF) method with an efficient iterative scheme, is one such method [34] in which one focuses on any local region and analyzes the effect of the rest of the system on it. Using this method, we first set up the force constant matrix, from analytical expressions of the partial second derivatives of the potential, in a layer-by-layer manner [35]. Because of the finite range of the atomic interactions, this matrix takes a block-tridiagonal form, allowing the Green's-function matrix corresponding to the local region of interest to the constructed following the procedure described in Ref. [34]. In this method of resolvent matrix, the calculation of the Green functions is reduced to a series of inversions and multiplications of matrices whose dimensions are usually smaller than the total number of degrees of freedom in the system. Another feature of the method is that one can focus on any specified locality in the system and calculate the corresponding properties. From the calculated Green function, normalized local vibrational densities of states are obtained using:

$$g(\nu^2) = -\frac{1}{3n\pi} \lim_{\varepsilon \to 0} \operatorname{Im} Tr[G(\nu^2 + i\varepsilon)] \qquad (3)$$

where n is the number of atoms in the chosen locality. The frequency dependent vibrational density of states N(ν) is related to $g(\nu^2)$ trough the equation:

$$N(\nu) = 2\nu g(\nu^2) \qquad (4)$$

## III.3. Thermodynamic Functions, Prefactors and Diffusion Coefficients

Once the local vibrational density of states is calculated, we can obtain all thermodynamic quantities for systems of interest from the partition function calculated within the harmonic approximation of lattice dynamics. The main quantity for our purposes here is the vibrational contribution to the free energy, which is given by the standard definition F = U − T S, where U is internal energy, S is entropy and T is temperature. Both U and S have contributions from atomic configurations and vibrations, such that F = $F^{conf}$ + $F^{vib}$. That is, for each atomic configuration of the system there is a specific vibrational contribution, which can be further written as $F^{vib}$ = $U^{vib}$ − $TS^{vib}$. In the harmonic approximation, vibrational part of the free energy is further given by:

$$F^{vib} = k_B T \int_0^\infty N(\nu) \ln\left(2\sinh\left(\frac{x}{2}\right)\right) d\nu \qquad (5)$$

while those contributions to the internal energy and entropy is obtained as:

$$U_{vib} = k_B T \int_0^\infty N(\nu)\left(\frac{1}{2}x + \frac{x}{e^x - 1}\right) d\nu \qquad (6)$$

$$S_{vib} = k_B \int_0^\infty N(\nu)\left(-\ln(1 - e^{-x}) + \frac{x}{e^x - 1}\right) d\nu \qquad (7)$$

Here $k_B$ is the Boltzman constant, T is the temperature, $x = h\nu/k_B T$ (h is the Plank constant) and $N(\nu)$ is the vibrational density of states (as a function of frequency $\nu$) which can be expanded as $N(\nu) = \sum_l n_l(\nu)$, where $n_l(\nu)$ is the local density of states (LDOS) of atoms in layer $l$.

For an isolated atom migrating on a surface, the diffusion coefficient D may be obtained from the Einstein relation for a random walk, D=<$\Delta r^2$>/2$\alpha$t, where <$\Delta r^2$>=$Nd^2$ is the mean-square displacement of the diffusing particle during the time period t, $\alpha$ is the dimensionality of the motion, and d is the jump distance. The number of jumps N is the product of the time period and a hopping rate $\Gamma$, which for thermally activated diffusion may be expressed according to transition state theory [36] as;

$$\Gamma = \frac{k_B T}{h} \exp(\frac{-\Delta F}{k_B T}) \qquad (8)$$

where, $\Delta F$ is the difference in the Helmholtz free energy between the two states where the system is at the minimum energy and at the saddle point, here $F$= E+U-TS, with E, U and S being the potential energy, the internal energy and the entropy of the system, respectively. The essential feature of this equation is the dependence of $\Gamma$ on the free energy. Using the thermodynamic quantities in equations 5, 6 and 7, we re-write the diffusion coefficient D as:

$$D = D_0(T) \exp\left(\frac{-\Delta E}{k_B T}\right) \qquad (9)$$

$$D_0(T) = \frac{k_B T}{h} \frac{nd^2}{2\alpha} \exp\left(\frac{\Delta S_{vib}}{k_B}\right) \exp\left(\frac{-\Delta U_{vib}}{k_B T}\right) \qquad (10)$$

The values of n (the number of equivalent jumps), d and $\alpha$ for each process (P1-P13), as previously defined in the literature [17,25,37], are given in Table 2.

**IV. Result and Discussion**

In this section, we present our results and compare them with existing experimental data and other theoretical values whenever available. The diffusion activation energies for the processes P1-P13 in Fig 1 and 2 are discussed first, followed by the specifics of vibrational densities of states for selected cases, and the contribution of vibrational dynamics to the diffusion characteristics. Note that the activation energy barriers for several processes and the diffusion prefactors for some processes have already been reported in the literature. We do find the need to present them here for the comparison and completeness necessary to make our point.

**IV.1 Activation Barriers**

Our results for the calculated activation barriers for adatom hopping processes (P1-P6 in Fig 1) on Cu(100) are presented in Table I and compared to the available experimental and theoretical values. Clearly, the activation energy associated with the process P1 (0.505 eV) is in quite good agreement with results of previous theoretical studies using, ab-initio methods and EAM potentials (with different parameterizations Voter and Chen (VC) [38], Adams, Foils and Wolfer (AFW) [39] and Foiles, Baskes and Daw (FBD) [29]). Compared to the experimental results, however, we note that the theoretical results overestimate this energy barrier. While He scattering measurements find an activation barrier of 0.28±0.06 eV [40], the low energy ion scattering (LEIS) experiments report them to be 0.39±0.06 eV [36] and 0.36±0.06 eV [41].

From the Table I, we also get a comparison of the activation energy barriers for the cases when the adatom is near a step edge (P2-P6). We find the activation barriers for the adatom to diffuse away from the step (P2) and along the step edge (P4) to be, respectively, 0.483 eV and 0.499 eV which are only slightly different from that for the adatom diffusing on a flat Cu(100) surface (P1). This reflects that the presence of the step does not noticeably perturb the energy landscape associated with these two processes. On the other hand, in agreement with previous theoretical results [27, 42], our calculated energy barrier for adatom jump to lower terrace from a step edge (P3) and away from a step edge (P5), are 0.79 eV and 0.847 eV, respectively, and are considerably higher. These results indicate the tendency of an adatom to stay close to step edges. At the same time, and again in agreement with previous findings [27, 42], the barrier for adatom diffusion along the step edge, at the lower terrace (P6), is relatively small at 0.265 eV. An adatom would thus prefer to diffuse along the step, to the lower terrace, rather than diffuse away from it. The issue, of course, is whether the prefactors are dramatically different. We come back to this point in section IV.3.

Turning now to the activation energy barriers for adatom hopping on diffusion Cu(110), we should note that geometric anisotropy of this surface leads to interesting variety in the processes. The diffusion activation energy along the open channel (P7) is found to be 0.230 eV which is similar to that of earlier work [27, 26]. The activation barrier for diffusion perpendicular to the open channel (P8) is found to be much larger, as expected, at 1.146 eV (1.15 eV in ref. [27]). The dense chains of atoms on Cu(110) are separated by a length larger than the nearest neighbor distance making the diffusion parallel to these chains (open channel) more facile than the direction perpendicular to this channel.

Now we turn to the case of an adatom near a step edge and consider its diffusion via hopping (P9-P13). Note that for these processes, we did not find any previous calculations, or experimental results to compare our results with. When the adatom is on the upper terrace, the diffusion activation barriers corresponding to jump over the step (P9), away from the step (P10) and along the step (P11) are 0.644 eV, 0.222 eV and 1.131 eV, respectively. Again, we find the presence of the step not to influence the energy landscape along and away from it since the barrier energy for P10 is very close to that of P7 and that of P11 close to that of P8. For the case of the adatom on the lower terrace, the activation barriers of the adatom are found to be 0.475 eV for P12 0.864 eV for P13, signifying decreased mobility along the step edge as compared to that away from it. This is also not surprising since the step edge atoms on this surface have a coordination of 6 and provide a kinked edge offering the adatom enhanced opportunities for bonding, as compared to the smother edge on the Cu(100) step discussed earlier.

**IV.2 Local Densities of States (LDOS)**

We present here results for only two cases (P7 and P8) on Cu(110) to illustrate the important characteristics of the vibrational dynamics of the adatom at both equilibrium configuration and saddle point sites. In Fig.3, we have plotted the x, y and z resolved local density of states for the adatom at the hollow site on Cu(110). The densities of states for the adatom at the saddle point, corresponding to P7 and P8 are shown on figures 4 and 5, respectively. From fig. 3, we note that the contribution of the x component of the density of state presents a dramatic softening of the low frequency end of the spectrum resulting from a substantial reduction of the force field along the x direction (open channel).

When the adatom is placed in the saddle point configuration, it is actually constrained along the diffusion direction and hence has only two degrees of freedom. From figures 4 and 5 one can note a common feature to the two cases, which is the appearance of high frequency peaks. In figure 4, the adatom is constrained in x-direction (P7), and the high frequency mode is found in the y component of the LDOS, reflecting the fact that the adatom is in the open channel with a strong bond (stiffening along y) with the atoms on the two surrounding chains. In the case of P8, (see figure 5), the adatom is constrained in y-direction and the high frequency peak is seen in the z component of the LDOS. Here, the bond with the atoms just below the adatom at the saddle point is stiffened along the z direction. These findings are consistent with what has been mentioned in earlier work using EAM [18,17,25,43].

**IV.3 Prefactors and Diffusion Coefficients**

In this section we discuss our calculated prefactors and diffusion coefficients and compare them to available experimental data and theoretical calculations at 300K and 600K. These results are presented in Table 2. For self-diffusion of an adatom on Cu(100) (P1) our calculated prefactor is found to be $7.29 \times 10^{-4}$ cm$^2$/s at 300K. This value is similar to those from previous theoretical work and experimental data as can be seen in Table 2. For the case of self-diffusion on Cu(110) terrace, our calculated prefactor values are $6.29 \times 10^{-4}$ cm$^2$/s for (P7), and $9.97 \times 10^{-4}$ cm$^2$/s for (P8). For process (P7), previous theoretical study [37] found values of $8 \times 10^{-4}$ cm$^2$/s and $4 \times 10^{-4}$ cm$^2$/s using two different types of EAM potentials. For diffusion perpendicular to the open channel (P8) the prefactors from the two types of potential were found to be $3.2 \times 10^{-3}$ cm$^2$/s and $2.7 \times 10^{-4}$ cm$^2$/s [37]. The interesting point to note is that almost all values of prefactors in Table 2 lie in the range $10^{-3}$ cm$^2$/s. In figure 6, we plot the prefactor versus activation energy barrier to determine if there is any correlation supporting the Meyer Neldel compensation rule [24]. We note that only a few cases follow the compensation rule (P7, P12, P9, P13, and P11) for Cu(110), for which the activation energies vary from 0.23 eV to 1.13 eV and the prefactor ranges between $6.29 \times 10^{-4}$ cm$^2$/s and $8.6 \times 10^{-3}$ cm$^2$/s). However, in general, this rule is not obeyed. Our conclusion is in qualitative agreement with that reported by an experimental study [44].

**V. Conclusion**

In this paper we have investigated the role of vibrational entropy on several diffusion processess on Cu(100) and Cu(110) terraces and near step edges. We have calculated diffusion barriers, prefactors, and diffusion coefficients along several diffusion processes for adatom hopping, using the harmonic approximation of lattice dynamics. On Cu(100) the prefactors are in the range from $5.53 \times 10^{-4}$ to $2.89 \times 10^{-3}$ cm$^2$/s at 300K, in good agreement with available experimental data and previous calculations. For adatom hopping on Cu(110) terraces and step edges, the calculated prefactors range from $6.29 \times 10^{-4}$ to $8.6 \times 10^{-3}$. We also find that the temperature has very little effect on these values when raising it to 600K. Though the Meyer-Nelded compensation rule is not found to be followed in general, most of the processes on Cu(110) studied here show a certain tendency of the compensation.

**ACKNOWLEDGEMENTS:** This work was supported by NSF grant INT-0244191 and NSF-TUBITAK international program under the Grant No. TBAG-U/59(102T210).

**REFERENCES**
**[1]** E. W. Muller, Z. Electrochem. **61,** 43 (1957).
**[2]** G. Ehrlich, F.G. Hudda, J. Chem. Phys. **44**. 1039 (1966).
**[3]** S.C. Wang and G. Ehrlich, Surf. Sci. **224**, L997 (1989);  S.C.Wang and G. Ehrlich, Phys. Rev. Lett. **67**, 2509 (1991).
**[4]**  T.T. Tsong, Phys. Rev. B **6**, 417 (1972); T.T. Tsong, R. J. Walko, Phys. Status Solid A **12**, 11 (1972).
**[5]**  T.T. Tsong, and G.L. Kellogg, Phys. Rev. B **12**, 1343 (1975).
**[6]**  G. Ayrault and G. Ehrlich, J. Chem. Phys. **60**, 281 (1974).
**[7]**  D.W. Bassett and P.R. Webber, Surf. Sci. **70**, 520 (1978); G.L. Kellogg Surf. Sci. **246**, 31-36, (1991).
**[8]**  D.W. Bassett, J. Phys. C (Solid State Physics) **9**, 2491 (1976).
**[9]**  R.T. Tung and W.R. Graham, Surf. Sci. **97**, 73 (1980); G.L. Kellogg Surf. Sci. **266**, 18-23, (1992).
**[10]**  D.W. Bassett and M.J. Parsley, J. Phys. D (Appl. Phys.) **2**, 13 (1969).
**[11]**  G.L. Kellogg, Surf. Sci. Rep. **21**, 1 (1994).
**[12]**  G. Boisvert, N. Mousseau, and L.L. Lewis, Phys. Rev. B **58**, 12667 (1998).
**[13]**  Z. Chvoj and M.C. Tringides, Phys. Rev. B **66**, 035419 (2002); K.R. Roosand and M.C. Tringides, Phys. Rev. Lett. **85**, 1480 (2000).
**[14]**   E. Kaxiras, and K.C. Pandey, Phys. Rev. B **47**, 1659 (1993).
**[15]**  L.B. Hansen, P. Stoltze, K. W. Jacobsen, and J. K. Norskov**,** Surf. Sci. **289**, 68 (1993).
**[16]**  G. H. Vineyard, J. Phys. Chem. Solids **3**, 121 (1957).
**[17]** U. Kurpick, A. Kara, T. S. Rahman, Phys. Rev. Lett. **78,** 1086 (1997).

**[18]** U. Kurpick, T.S. Rahman, Surf. Sci. **383**, 137 (1997); U. Kurpick, T.S. Rahman, Phys. Rev. B **57**, 2482 (1998)

**[19]** U. Kurpick, T.S. Rahman, Surf. Sci. **15-21,** 427-428 (1999).

**[20]** U. Kurpick, T. S. Rahman, Phys. Rev. B. **59,** 11014 (1999).

**[21]** H. Eyring, J. Chem. Phys. **3**, 107 (1935).

**[22]** W. Meyer and H. Neldel, Z. Tech. Phys. **12,** 588 (1937)**.**

**[23]** A. Yelon and B. Movaghar, Phys. Rev. Lett. **65,** 618 (1990); A. Yelon, B. Movaghar, and H. M. Branz, Phys. Rev. B **46**, 12 244 (1992).

**[24]** G. Boisvert and J. L. Lewis, and A. Yelon, Phys. Rev. Lett. **75,** 469 (1995).

**[25]** U. Kurpick, Phys. Rev. B **64,** 075418 (2001).

**[26]** F. Montalenti and R. Ferrando, Phys. Rev. B. **59,** 5881 (1999).

**[27]** M. Karimi, T. Tomkovski, G. Vidali, and O. Biham, Phys. Rev. B. **52,** 5364 (1995).

**[28]** W. Press, S. Teukolsky, W. Vetterling, and B. Flannery 1992 Numerical Recipes in Fortran (Cambridge: Cambridge University Press).

**[29]** S. M. Foiles, M. I. Baskes, M. S. Daw, Phys. Rev. B **33,** 7983 (1996); M. S. Daw, S. M. Foiles, and M. I. Baskes, Mater. Sci. Rep. **9,** 251 (1993).

**[30]** H. Jonsson, G. Mills and K. W. Jacobsen, Classical and Quantum Dynamics in Condensed Phase Simulations ed. by B. J. Berne World Scientific, Singapore, (1998).

**[31]** T. S. Rahman, A. Kara, A. Karim and A. Al-Rawi, Collective Diffusion on Surfaces: Correlation Effects and Adatom Interactions, ed. By M.C Tringides and Z. Chvoj (Kluwer 2001).

**[32]** R. E. Allen, G. P. Alldredge, and de F. W. Wette Phys. Rev. B **4,** 1648 (1971).

**[33]** R. Haydock, V. Heine and M J. Kelly, J. Phys. C: Solid State Phys. **5,** 2845 (1972).

**[34]** S. Y. Wu, J. Cocks and C. S. Jayanthi, Phys. Rev. B **49,** 7957 (1994).

**[35]** A. Kara, C. S. Jayanthi, S. Y. Wu and F. Ercolessi, Phys. Rev. Lett. **72,** 2223 (1994) A. Kara, C. S. Jayanthi, S. Y. Wu and F. Ercolessi, Phys. Rev. B **51**, 17046 (1995).

**[36]** M. Breeman and D. O. Boerma, Surf. Sci. **269/270,** 224 (1992).

**[37]** C. L. Liu, J. M. Cohen, J. B. Adams, A. F. Voter, Surf. Sci. **253,** 334-344 (1991).

**[38]** A.F. Voter and S.P. Chen, Materials Research Society Symposium Proceedings (1987).

**[39]** J. B. Adams, S.M. Foiles and W.G. Wolfer, J. Mater. Res. **4**, 102 (1989).

**[40]** H. -J. Ernst, F. Fabre and J. Lapujoulade, Phys. Rev. B. **46,** 1929 (1992).

**[41]** H. Durr, J. F. Wendelken and J. K. Zuo, Surf. Sci. **328,** L527 (1995).

**[42]** Z. –J. Tian, T. S. Rahman, Phys. Rev. B. **47,** 9751 (1993)**.**

**[43]** U. Kurpick, Phys. Rev. B **63,** 045409 (2001).

**[44]** S. C. Wang and G. Ehrlich, Surf. Sci. **206,** 451 (1988).

**[45]** J. J. Miguel et al, Surf. Sci., **189/190,** 1062 (1987).

**[46]** G. Boisvert and J. L. Lewis, Phys. Rev. B. **56,** 7643 (1997).

**[47]** L. Hansen, P. Stoltze, K. W. Jacobsen, and J. K. Norksov, Phys. Rev. B. **44,** 6523 (1991).

**FIGURE CAPTIONS**

**Fig.1** Diffusion of an adatom via hopping on fcc(100) on the terrace and at the step. The processes are described in the text.

**Fig.2.** Same as figure 1 for fcc(110).

**Fig. 3.** LDOS for the adatom on Cu(110) in its minimum energy site

**Fig. 4.** LDOS for the adatom on Cu(110) at the saddle point for P7.

**Fig. 5.** Same as figure 4 for P8.

**Fig. 6**. Activation barriers versus prefactors for each processes on (100) and (110).

**Table 1.** Activation Barriers. EX: Exchange Process

| Processes | | Activation Barriers (eV) (present work) | Activation Barriers (eV) (Available data) |
|---|---|---|---|
| **P1** | Cu(100) | 0.505 | 0.48 [17]<br>0.40 [45]<br>0.28±0.06 [40]<br>0.39±0.06 [36]<br>0.36±0.06 [41]<br>0.44 [20]<br>0.52±0.05 [46]<br>0.48 [27]<br>0.44 [25]<br>0.38 [37]-AFW<br>0.53 [37]-VC<br>0.51 [19]<br>0.43 [47]<br>0.49 [42]<br>0.53 [26]<br>0.79 [EX-26] |
| **P2** Cu(100)/with-up-step/[away from the step] | | 0.483 | |
| **P3** Cu(100)/with-up-step/[over the step] | | 0.79 | 0.77 [27,26]<br>0.51 [EX-26] |
| **P4** Cu(100)/with-up-step/[along the step] | | 0.499 | |
| **P5** Cu(100)/with-down-step/[away from the step] | | 0.847 | 0.84 [27]<br>0.83 [42] |
| **P6** Cu(100)/with-down-step/[along the step] | | 0.265 | 0.25 [27]<br>0.26 [42] |
| **P7** Cu(110)/[along open channel] | | 0.230 | 0.24 [27]<br>0.25 [25]<br>0.23 [37]-AFW<br>0.28 [37]-VC<br>0.23 [26]<br>0.87 [EX-27] |
| **P8** Cu(110)/[perpendicular to the open channel] | | 1.146 | 1.15 [27] |
| **P9** Cu(110)/with-up-step/[over the step] | | 0.644 | |
| **P10** Cu(110)/with-up-step/[away from the step] | | 0.222 | |
| **P11** Cu(110)/with-up-step/[along the step] | | 1.131 | |
| **P12** Cu(110)/with-down-step/[away from the step] | | 0.475 | |
| **P13** Cu(110)/with-down-step/[along the step] | | 0.864 | |

**Table 2.** Prefactors and diffusion coefficients

| Processes | n, d, α | $D_0(T)$ (cm$^2$/s) 300K | $D_0(T)$ (cm$^2$/s) 600K | $D(T)$ (cm$^2$/s) 300K | $D(T)$ (cm$^2$/s) 600K | $\Delta F_{vib}$ (meV) 300K | $\Delta F_{vib}$ (meV) 600K |
|---|---|---|---|---|---|---|---|
| **P1** | n =4<br>d=2.556<br>α =2 | 7.29x10$^{-4}$<br>8.7x10$^{-4}$ [19]<br>3.2x10$^{-4}$ [17]<br>2.5x10$^{-3}$ [25]<br>1.2x10$^{-3}$ [37]<br>5.2x10$^{-3}$ [37] | 7.43x10$^{-4}$ | 2.39x10$^{-12}$ | 4.26x10$^{-8}$ | 44.4 | 123.4 |
| **P2** | n =1<br>d =2.556<br>α =1 | 5.53x10$^{-4}$ | 5.64x10$^{-4}$ | 2.54x10$^{-17}$ | 1.2x10$^{-10}$ | 33.6 | 102.1 |
| **P3** | n =1<br>d =2.556<br>α =1 | 8x10$^{-4}$ | 7.97x10$^{-4}$ | 5.7x10$^{-12}$ | 6.72x10$^{-8}$ | 24.1 | 84.2 |
| **P4** | n =4<br>d =2.556<br>α =2 | 1.63x10$^{-3}$ | 1.62x10$^{-3}$ | 6.3x10$^{-12}$ | 1.0x10$^{-7}$ | 23.5 | 83.1 |
| **P5** | n =1<br>d =2.556<br>α =1 | 5.63x10$^{-4}$ | 5.73x10$^{-4}$ | 3.32x10$^{-18}$ | 4.4x10$^{-11}$ | 33.2 | 101.3 |
| **P6** | n =4<br>d =2.556<br>α =2 | 2.89 x10$^{-3}$ | 2.81x10$^{-3}$ | 1.02x10$^{-7}$ | 1.67x10$^{-5}$ | 8.78 | 54.9 |
| **P7** | n =2<br>d =2.556<br>α =1 | 6.29x10$^{-4}$<br>1.1x10$^{-3}$ 3 [25]<br>8x10$^{-4}$ [37]<br>4.4x10$^{-4}$ [37] | 6.39x10$^{-4}$ | 8.61x10$^{-8}$ | 7.48x10$^{-6}$ | 48.2 | 131.5 |
| **P8** | n =2<br>d =3.615<br>α =1 | 9.97x10$^{-4}$ | 1.1x10$^{-3}$ | 5.57x10$^{-23}$ | 2.4x10$^{-13}$ | 54.4 | 144.1 |
| **P9** | n =1<br>d =2.556<br>α =1 | 2.18x10$^{-3}$ | 2.12x10$^{-3}$ | 3.31x10$^{-14}$ | 8.25 x10$^{-9}$ | 0.00 | 33.7 |
| **P10** | n =1<br>d =2.556<br>α =1 | 1.32x10$^{-3}$ | 1.33x10$^{-3}$ | 2.46x10$^{-7}$ | 1.81x10$^{-5}$ | 11.1 | 58.0 |
| **P11** | n =2<br>d =3.615<br>α =1 | 8.6x10$^{-3}$ | 8.4x10$^{-3}$ | 8.55x10$^{-22}$ | 2.65x10$^{-12}$ | 0.00 | 34.6 |
| **P12** | n =1<br>d =2.556<br>α =1 | 1.09x10$^{-3}$ | 1.10x10$^{-3}$ | 1.15x10$^{-11}$ | 1.12x10$^{-7}$ | 15.9 | 67.6 |
| **P13** | n =2<br>d =3.615<br>α =1 | 5.74x10$^{-3}$ | 5.72x10$^{-3}$ | 1.75x10$^{-16}$ | 3.2x10$^{-10}$ | 9.09 | 54.2 |

**Figure 1**

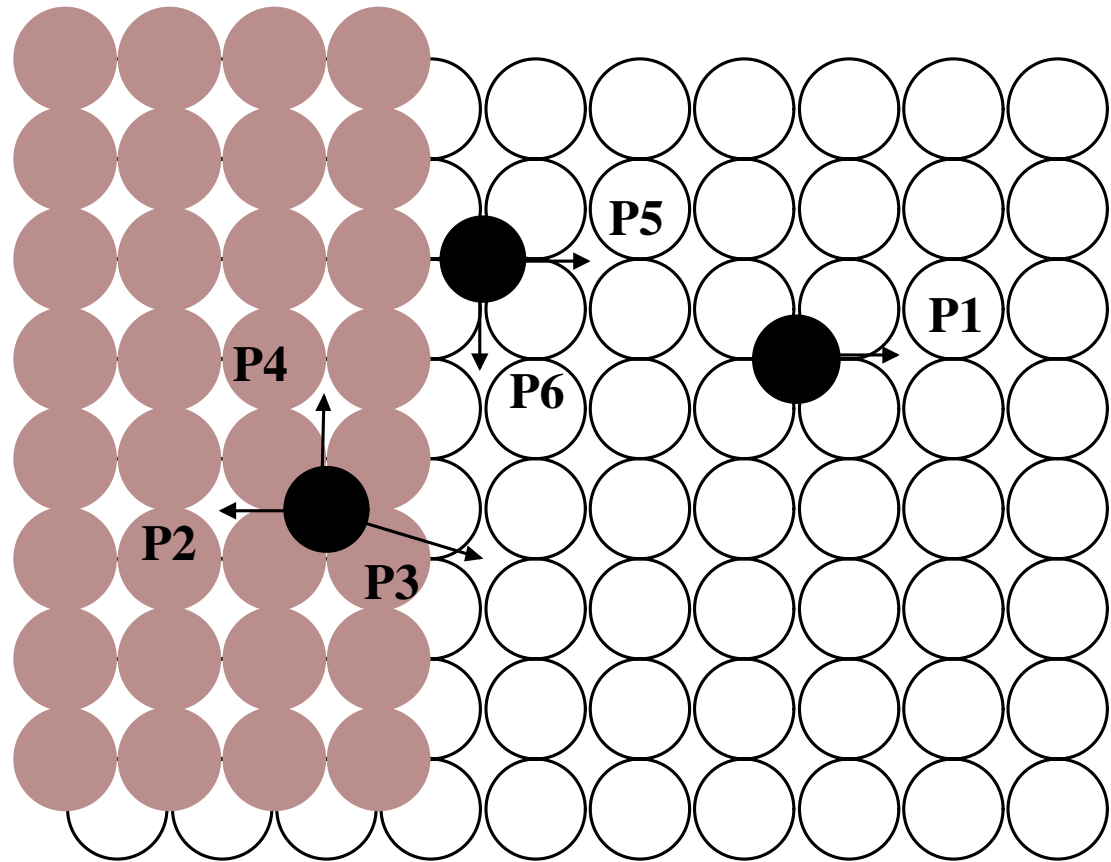

**Figure 2**

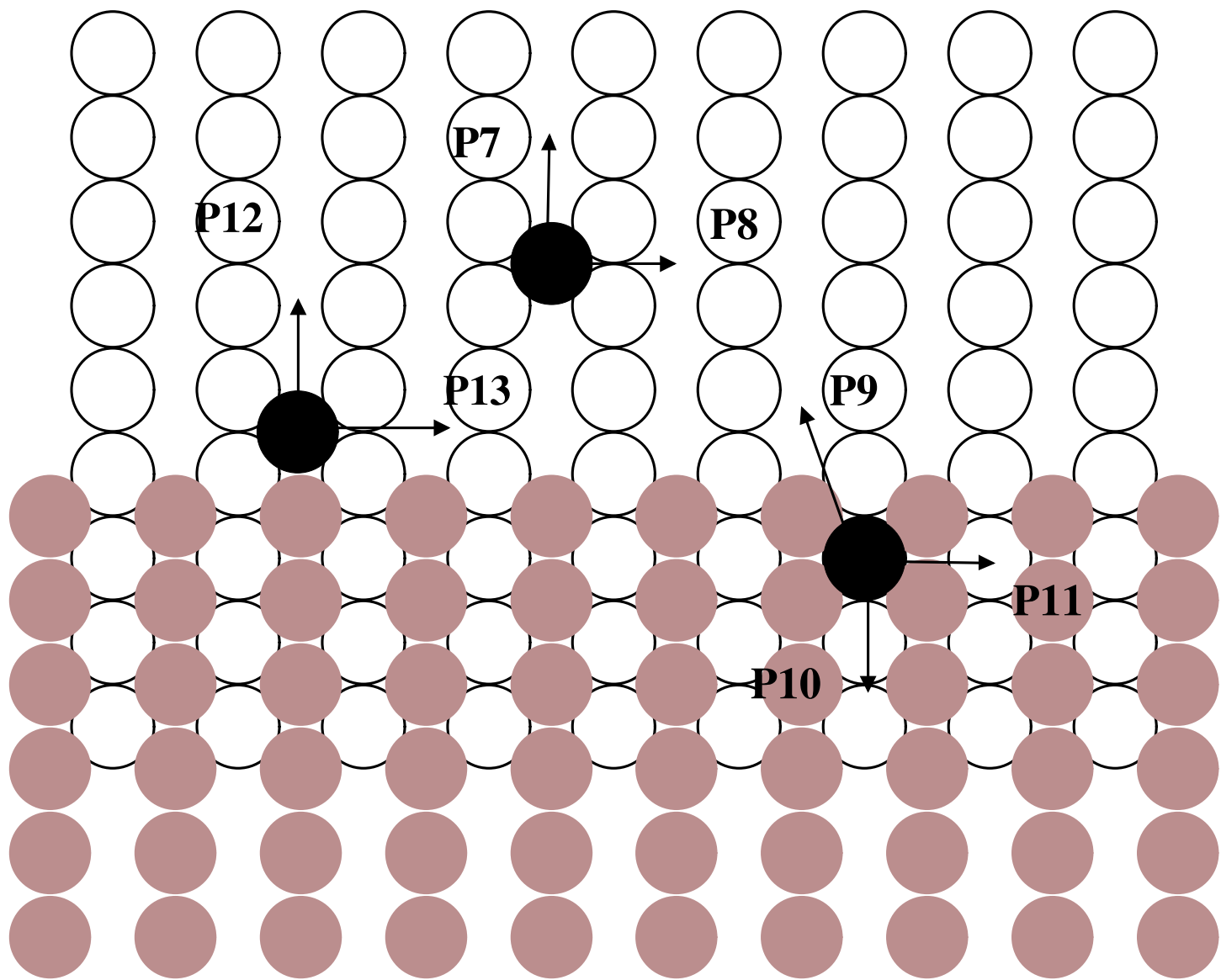

**Figure 3**

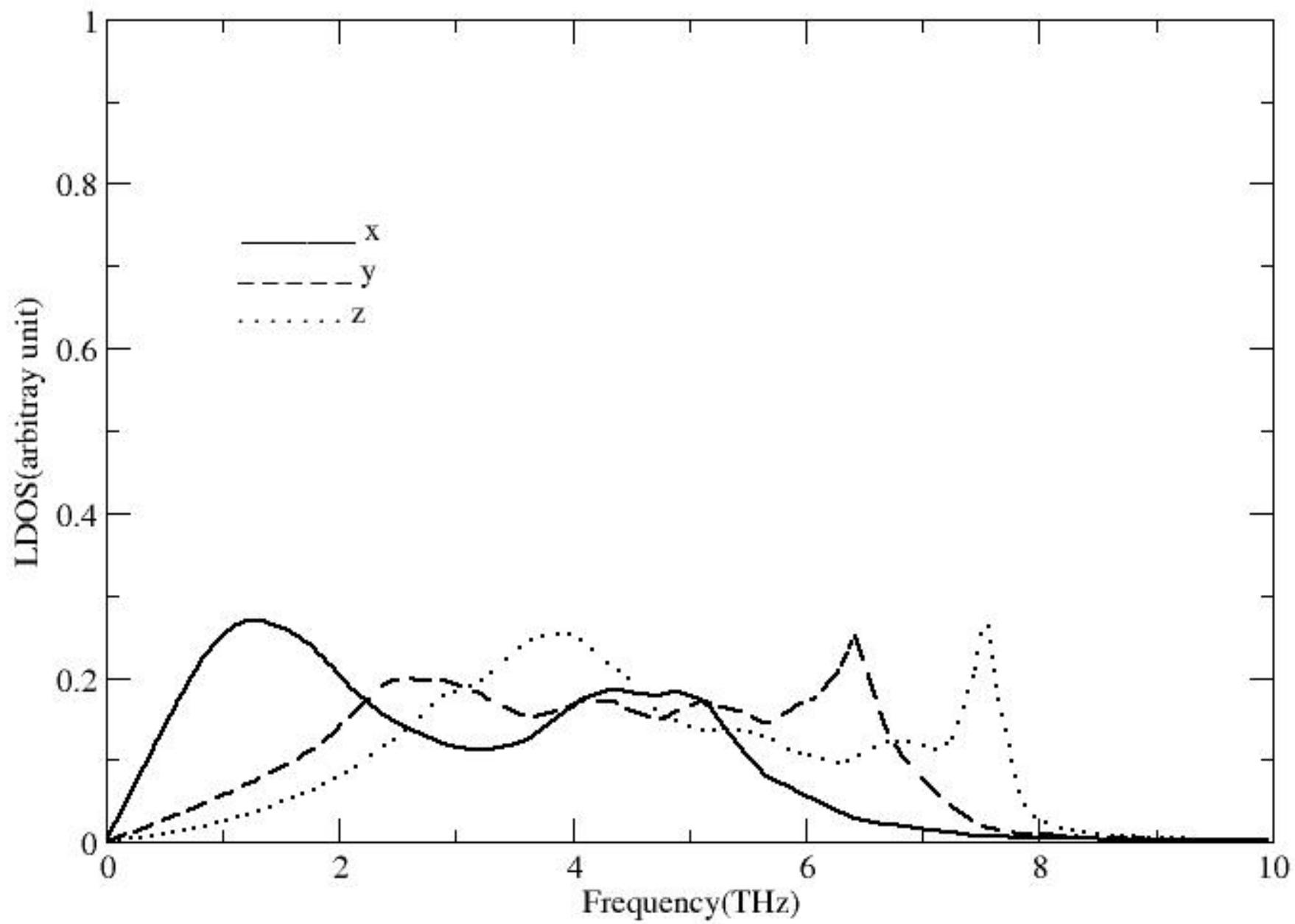

**Figure 4**

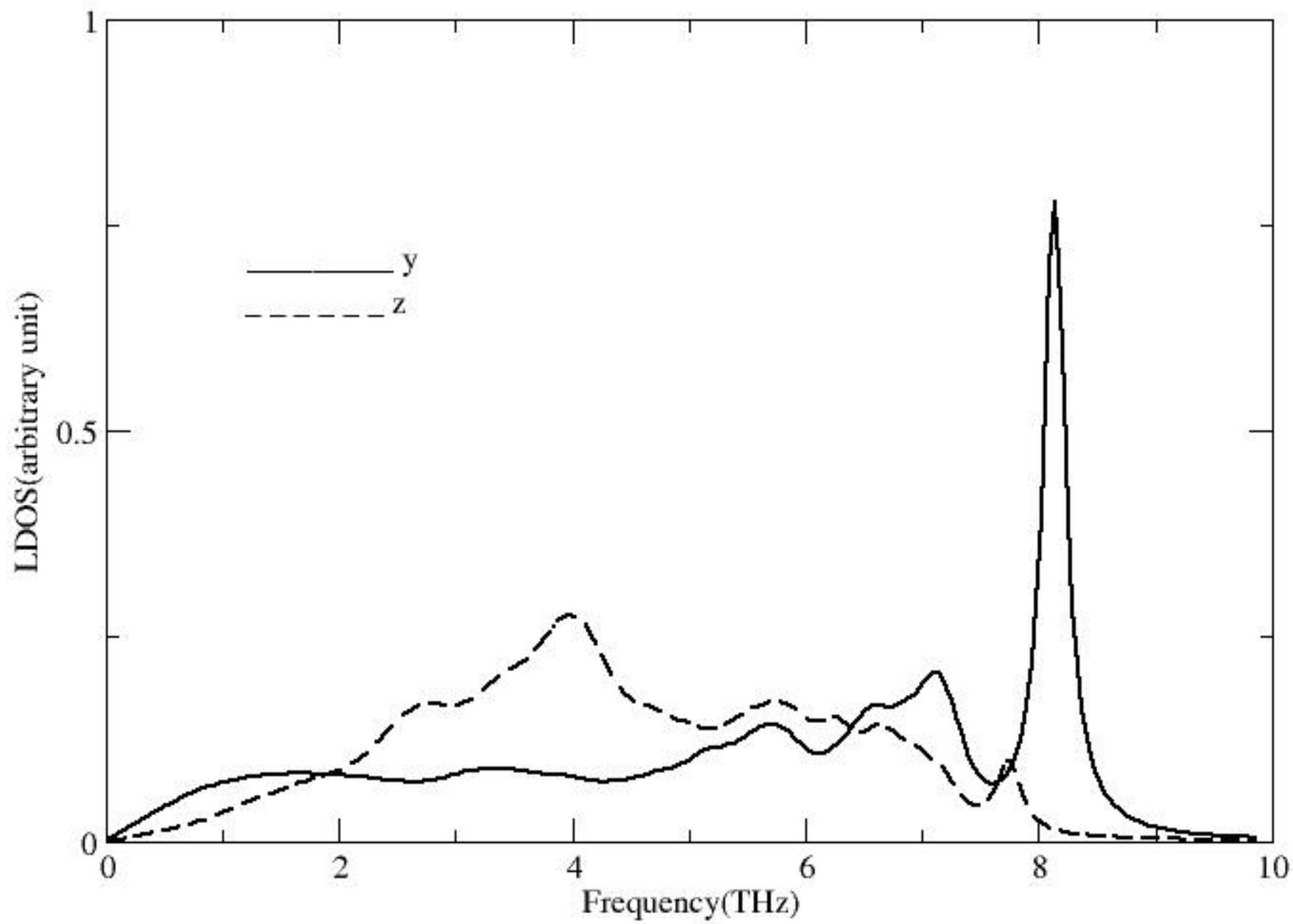

**Figure 5**

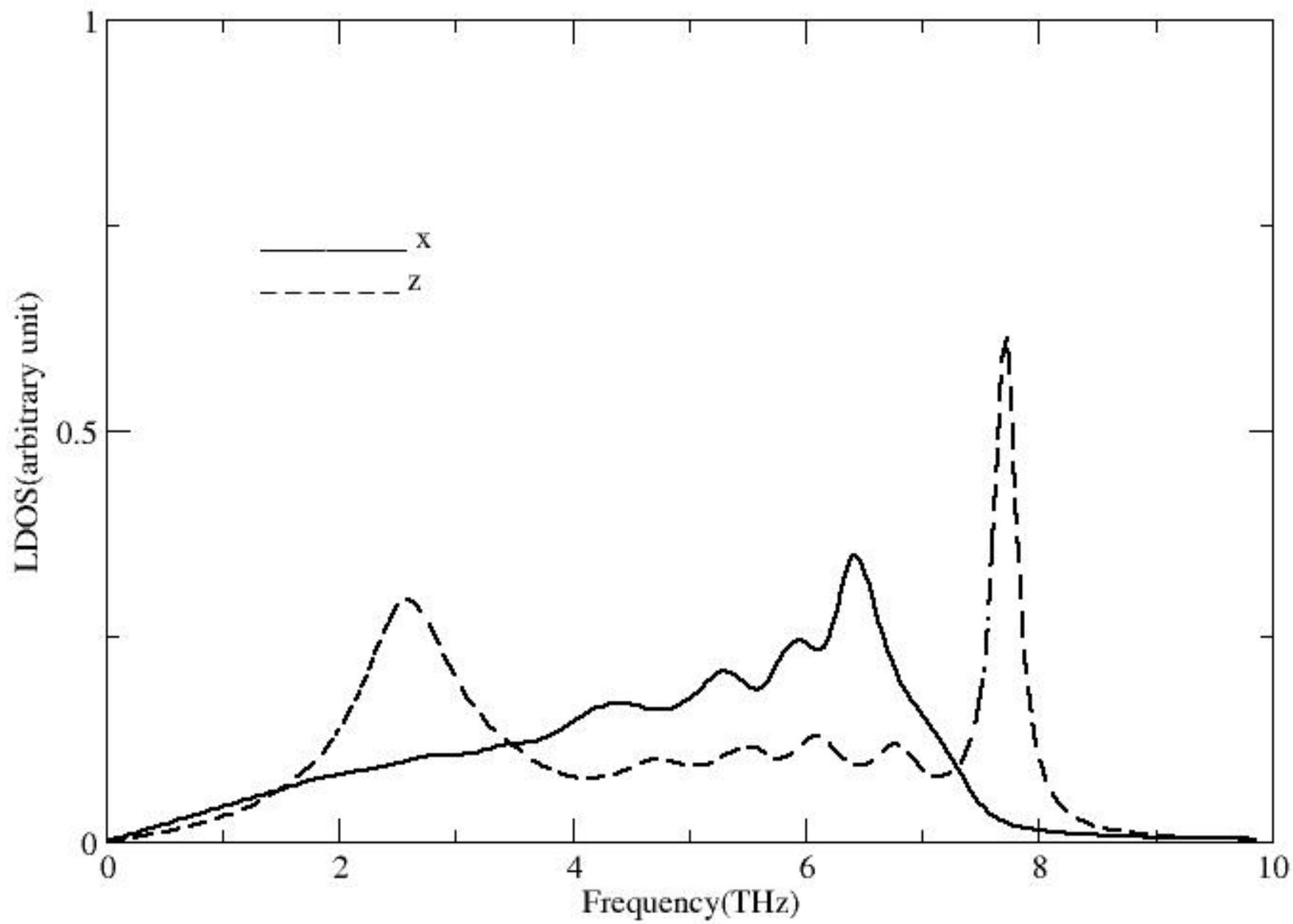

**Figure 6**

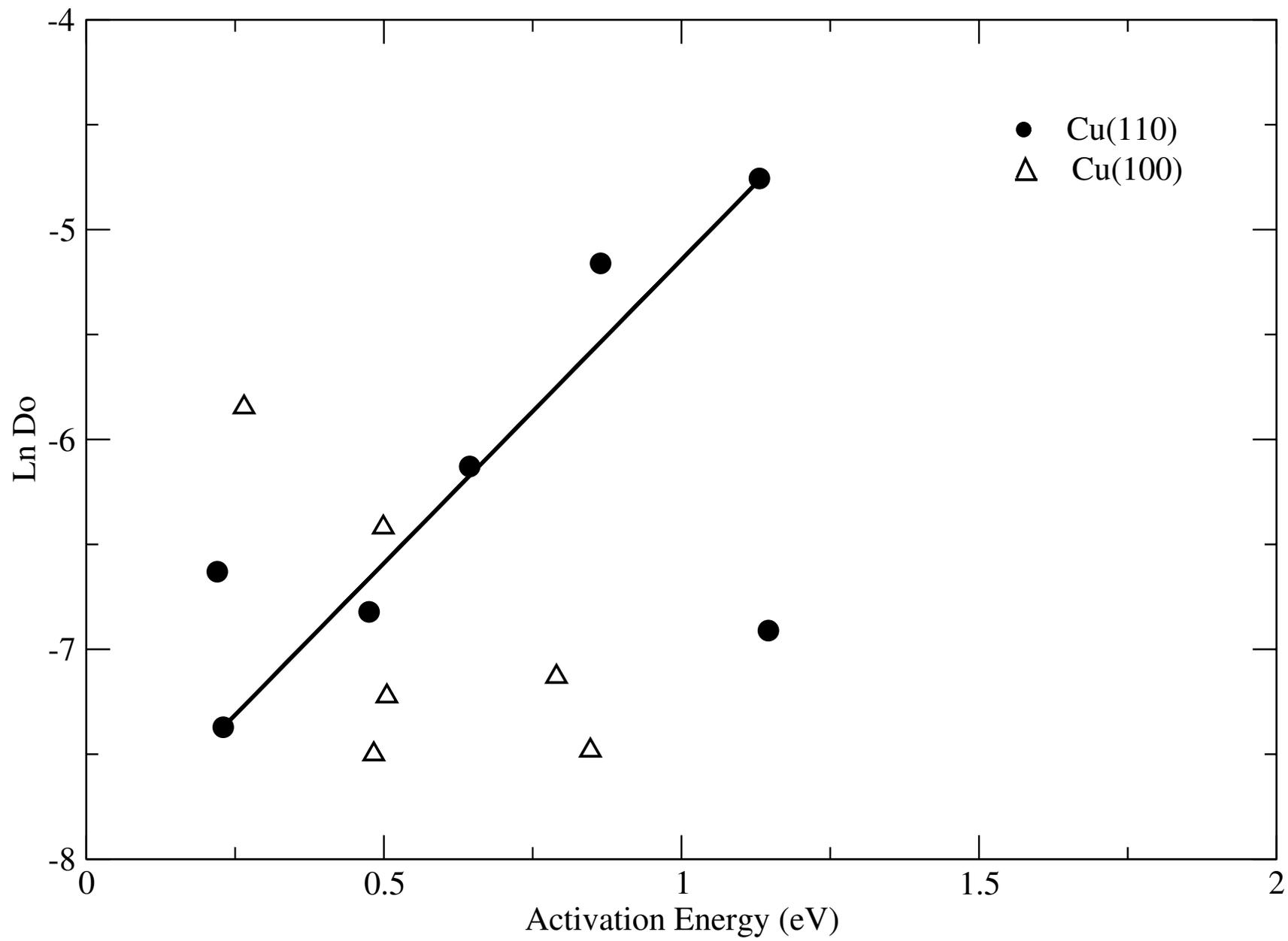